\documentclass[11pt]{article}

\usepackage{graphicx,amsmath,amsfonts,amssymb,dcolumn,amsthm,lipsum}
\usepackage[font=small]{caption}
\usepackage[labelfont=bf,labelsep=quad,justification=justified]{caption}
\begin{document}

\title{\textbf{Predicting Planck scale and Newtonian constant from a Yang-Mills gauge theory: 1 and 2-loops estimates}}
\author{\textbf{Rodrigo F.~Sobreiro$^1$}\thanks{sobreiro@if.uff.br} \ and \textbf{Anderson A.~Tomaz$^{1,2}$}\thanks{tomaz@cbpf.br}\\\\\
$^1$\textit{{\small UFF $-$ Universidade Federal Fluminense,}}\\
\textit{{\small Instituto de F\'{\i}sica, Campus da Praia Vermelha,}}\\
\textit{{\small Avenida General Milton Tavares de Souza, s/n, 24210-346,}}\\
\textit{{\small Niter\'oi, RJ, Brazil.}}\\\\
$^2$\textit{{\small CBPF $-$ Centro Brasileiro de Pesquisas F\'isicas,}}\\
\textit{{\small Rua Dr. Xavier Sigaud, 150 , Urca, 22290-180}}\\
\textit{{\small Rio de Janeiro, RJ, Brazil}}
}
\date{}
\maketitle

\begin{abstract}
Recently, a model for an emergent gravity based on $SO(5)$ Yang-Mills action in Euclidian 4-dimensional spacetime was proposed. In this work we provide some 1 and 2-loops computations and show that the model can accommodate suitable predicting values for the Newtonian constant. Moreover, it is shown that the typical scale of the expected transition between the quantum and the geometrodynamical theory is consistent with Planck scale. We also provide a discussion on the cosmological constant problem.
\end{abstract}

\section{Introduction}\label{INTRO}

Quantization of the gravitational field is one of most import problems in Physics since the beginning of the 20th century. The long pursuit of a theory of quantum gravity have generated a variety of theoretical proposals to describe the quantum sector of gravity, see for instance Loop Quantum Gravity \cite{Ashtekar:2004eh,Rovelli:2014ssa}, Higher Derivatives Quantum Gravity \cite{Asorey:1996hz,Buchbinder:1992rb}, Causal Sets \cite{Henson:2006kf}, Causal Dynamical Triangulations \cite{Ambjorn:2012jv}, String Theory \cite{Witten:1998qj,Seiberg:1999vs}, Asymptotic Safety \cite{Hawking:1979ig,Reuter:2012id}, Emergent Gravities \cite{Buchbinder:1992rb,Barcelo:2001tb}, Ho\v{r}ava-Lifshitz gravity \cite{Horava:2009uw}, the Nojiri-Odintsov-Ho\v{r}ava-Lifshitz instability free gravity \cite{Nojiri:2009th,Cognola:2016gjy}, Topological Gauge Theories \cite{Baez:1999sr} and so on. Each one of these theories carries its own set of advantages and disadvantages. On the other hand, gauge theories are relentless in describing the high energy regime of particle physics \cite{Itzykson:1980rh,Cheng:1985bj,Agashe:2014kda}. Hence, one can question if gravity could also be described by a gauge theory in its high energy sector. In fact, since the seminal papers \cite{Utiyama:1956sy,Kibble:1961ba,Sciama:1964wt} about gauge theoretical descriptions of gravity, it is known that gravity can be, at least, dressed as a gauge theory for the local isometries of spacetime. See also \cite{Trautman:1981fd,Mardones:1990qc}. Although consistent with general relativity, these models also have problems with its quantization.

In \cite{Sobreiro:2011hb} it was proposed an induced gravity model from a pure Yang-Mills theory based on de Sitter-type groups. In this model, gravity emerges as an effective phenomenon originated by a genuine Yang-Mills action in flat space. The transition between the quantum gauge sector and the classical geometrical sector is mediated by a mass parameter, identified with the Gribov parameter \cite{Gribov:1977wm,Singer:1978dk,Zwanziger1989, Zwanziger:1989mf,Sobreiro:2005ec, Dudal:2005na,Baulieu:2008fy,Baulieu:2009xr, Dudal:2010cd,Dudal:2011gd,Capri:2012hh, Dudal:2012sb,Pereira:2013aza,Dudal:2013vha,Capri:2013naa, Capri:2015pja,Capri:2015ixa,Capri:2015nzw,Capri:2016aqq}. The combination of the running of this parameter and the running of the coupling parameter would provide a good scenario for an In\"on\"u-Wigner contraction \cite{Inonu:1953sp} of the gauge group, deforming it to a Poincar\'e-type group. Because, the original action is not invariant under the resulting group, the model actually suffers a dynamical symmetry breaking to Lorentz-type groups. At this point, with the help of the mass parameter, the gauge degrees of freedom are identified with geometrical objects, namely, the vierbein and spin-connection. At the same time the mass parameter and the coupling parameter are combined to generate the gravity Newtonian constant and an emergent gravity is realized. Moreover, a cosmological constant inherent to the model is also generated. See also \cite{Sobreiro:2012iv,Sobreiro:2012dp} for details.

The aim of the present work is to provide estimates for the emergent parameters of the model above discussed by applying the usual aparathus of quantum field theory (QFT). We concentrate our efforts at 1 and 2-loops computations. In particular, from the explicit expressions of the running coupling and the Gribov parameters, we are able to fit the actual value of Newtonian constant and to obtain a renormalization group cut-off very close to the Planck scale, as expected to be the transition scale from quantum to classical gravity.

The cosmological constant is an essential point, which can be related with the accelerated expansion of the Universe. Observational data and quantum field theory prediction for the cosmological constant strongly disagree in numerical values \cite{Weinberg:1988cp,Agashe:2014kda}. Following \cite{Shapiro:2006qx,Shapiro:2009dh}, we can expect that the cosmological parameter generated by the model should combine with  the value found in theoretical calculations by quantum field theory in a way that the effective final value fits the observational data. It is worth mention that the cosmological parameter generated by the model is related with the Gribov parameter \cite{Sobreiro:2011hb}. In \cite{Assimos:2013eua}, a preliminar estimative for these running parameters at 1-loop approximation was scratched. From this reasonable starting, we develop here a refinement on early predictions and we show a numerical improvements at 1-loop calculations. Further, we improove the techniques up to 2-loops estimates. Hence, we present a best estimative for the Gribov parameter in order to fit the model with a suitable emergent gravity.

 This work is organized as follows: In Sect.~\ref{EFGRAV} we resume some concepts and ideas about our effective gravity model. In Sect.~\ref{1LOOP}, the first results that we obtained for 1-loop estimates. In Sect.~\ref{2LOOP}, we present the main calculations results for running parameters at 2-loops are performed. In last Sect.~\ref{FINAL}, we discuss shortly our results and some perspectives will be cast.
 
\newpage

\section{Effective gravity from a gauge theory}\label{EFGRAV}

In \cite{Sobreiro:2011hb}, a quantum gravity theory was constructed based on an analogy with quantum chromodynamics, see also \cite{Sobreiro:2012iv,Sobreiro:2012dp}. In this section we will briefly discuss the main ideas, definitions and conventions behind this model\footnote{Even though most of the material in this section can be found in previous articles \cite{Sobreiro:2011hb,Sobreiro:2012iv,Sobreiro:2012dp}, some new aspects are not fully discussed there.}. 

The starting action is the Yang-Mills action,
\begin{equation}
S_{\mathrm{YM}}=\frac{1}{2}\int{F_A}^B\ast{F_B}^A\;,\label{ym1}
\end{equation}
where ${F^A}_B$ are the field strength 2-form, $F = dY + \kappa YY$, $d$ is the exterior derivative, $\kappa$ is the coupling parameter and $Y$ is the gauge connection 1-form, \emph{i.e.}, the fundamental field in the adjoint representation. The Hodge dual operator in $4$-dimensional Euclidian spacetime is denoted by $\ast$. The action Eq.~\eqref{ym1} is invariant under $SO(5)$ gauge transformations, $Y\longmapsto U^{-1}\left(1/\kappa \mathrm{d}+Y\right)U$, with $U\in SO(5)$. The infinitesimal version of the gauge transformation is
\begin{equation}
Y\longmapsto Y+\nabla\alpha\;,\label{gt1}
\end{equation}
where $\nabla=\mathrm{d}+\kappa Y$ is the full covariant derivative and $\alpha$ is the infinitesimal gauge parameter.

It is possible to decompose the gauge group according to $SO(5)=SO(4)\otimes S(4)$, where $SO(4)$ is the stability group $S(4)$ is the symmetric coset space. Thus, defining $J^{5a}\equiv J^a$, the gauge field is also decomposed,
\begin{equation}
Y={Y^A}_{\;B} {J_A}^{\;B} = {A^a}_{\;b}{J_a}^{\;b}+\theta^aJ_a\;,\label{connec1}
\end{equation}
where capital Latin indices $A,B,\dots$ run as $\{5,0,1,2,3\}$ and the small Latin indices $a,b,\dots$ vary as $\{0,1,2,3\}$. The decomposed field strength reads
\begin{equation} \label{Fdec}
 F = F^A_{\phantom{A}B} J_A^{\phantom{A}B} = \left( \Omega_{\ b}^{a} - \dfrac{\kappa}{4}\theta^{a}\theta{}_{b}\right)J_{a}^{\ b} + K^{a}J_{a},
\end{equation}
where $ \Omega_{\ b}^{a}=\mathrm{d}A_{\ b}^{a}+\kappa A_{\ c}^{a}A_{\ b}^{c} $ e $ K^{a} = \mathrm{d}\theta^{a}+\kappa A_{\ b}^{a}\theta^{b} $. Thus, it is a simple task to find that the Yang-Mills action Eq.~\eqref{ym1} can be rewritten as
\begin{equation} \label{ym2}
 S_{YM} = \frac{1}{2}\int \left\{ \Omega_{\ b}^{a}\ast\Omega_{a}^{\ b} + \frac{1}{2}K^{a}\ast K_{a} - \frac{\kappa}{2}\Omega_{\ b}^{a}\ast\left(\theta_{a}\theta^{b}\right)+\frac{\kappa^{2}}{16}\theta^{a}\theta_{b}\ast\left(\theta_{a}\theta^{b}\right)\right\}.
\end{equation}

Before we advance to next stage of the model, let us quickly point out some important aspects of Yang-Mills theories and their analogy with a possible quantum gravity model. To start with, Yang-Mills theories present two very important properties, namely, renormalizability and asymptotic freedom \cite{Itzykson:1980rh}. The Yang-Mills action is, in fact, renormalizable, at least to all orders in perturbation theory \cite{Piguet:1995er} which means that it is stable at quantum level. In this context, the so called BRST symmetry has a fundamental hole. Asymptotic freedom \cite{Gross:1973id,Politzer:1973fx}, on the other hand, means that, at high energies, the coupling parameter is very small and we can use perturbation theory in our favor. However, as the energy decreases, the coupling parameter increases and the theory becomes highly non-perturbative. In this regime, the so called Gribov ambiguities problem takes place \cite{Gribov:1977wm,Singer:1978dk}. Essentially, the gauge fixing\footnote{Although we did not specify the gauge fixing constraint, the Gribov problem is a pathological issue plaguing all covariant gauges \cite{Singer:1978dk}. Nevertheless, we can antecipate for the reader that we will employ the Landau gauge fixing in this entire work.} is not strong enough to eliminate all spurious degrees of freedom from the Faddeev-Popov path integral; a residual gauge symmetry survives the Faddeev-Popov procedure. The elimination of the Gribov ambiguities is not entirely understood, however, it is known that a mass parameter is required and a soft BRST symmetry breaking associated with this parameter appears, see, for instance, \cite{Dudal:2005na,Baulieu:2008fy,Baulieu:2009xr,Dudal:2011gd, Pereira:2013aza,Zwanziger:1992qr,Maggiore:1993wq}.  This parameter is known as Gribov parameter $\gamma$, and it is fixed through minimization of the quantum action, $\delta\Sigma/\delta\gamma^2=0$, the so called gap equation. The action that describes the improved theory (free of infinitesimal ambiguities) is known as Gribov-Zwanziger action \cite{Zwanziger:1992qr} and has a more refined version \cite{Zwanziger1989,Dudal:2011gd} by taking into account a few dimension two operators and their condensation effects.

It is clear that the field $\theta$ has the same degrees of freedom that a soldering form in spacetime manifold (the vierbein). However, the field $\theta$ carries UV dimension 1 while the vierbein is dimensionless. The presence of a mass scale is then very important to identify the field $\theta$ with an effective soldering form. We will show in the next sections that the Gribov parameter is a very good candidate for this purpose. The next step is to perform the rescalings
\begin{eqnarray}\label{rescale}
A &\rightarrow & \dfrac{1}{\kappa}A~,\nonumber\\
\theta &\rightarrow & \dfrac{\gamma}{\kappa}\theta~,
\end{eqnarray}
at the action Eq.~\eqref{ym2}, achieving
\begin{equation}
S=\frac{1}{2\kappa^2}\int\left[\overline{\Omega}^a_{\phantom{a}b}{*}\overline{\Omega}_a^{\phantom{a}b}+\frac{\gamma^2}{2}\overline{K}^a{*}\overline{K}_a-\frac{\gamma^2}{2}\overline{\Omega}^a_{\phantom{a}b}{*}(\theta_a\theta^b)+\frac{\gamma^4}{16}\theta^a\theta_b{*}(\theta_a\theta^b)\right]\;,\label{ym3}
\end{equation}
where \small{$\overline{\Omega}^a_{\phantom{a}b}=\mathrm{d} {A}^a_{\phantom{a}b}+ {A}^a_{\phantom{a}c} {A}^c_{\phantom{c}b}$}, \small{$\overline{K}^a=\mathrm{D}\theta^a$} and the covariant derivative is now \small{$\mathrm{D}=\mathrm{d}+A$}.

The transition from the action \eqref{ym3} to a gravity action is performed by studying the running behavior of the quantity $\gamma/\kappa$. It is expected \cite{Sobreiro:2011hb} that this quantity vanishes for a specific energy scale. This property induces an In\"on\"u-Wigner contraction $SO(5)\longmapsto ISO(4)$ \cite{Inonu:1953sp}. However, since the action \eqref{ym3} is not invariant under $ISO(4)$ gauge transformations, the theory actually suffers a symmetry breaking to the stability group $SO(4)$. The broken theory is ready to be rewritten as a gravity theory. The map (see \cite{Sobreiro:2011hb}) is a simple identification of the gauge fields with geometric effective entities according to $\delta_\mathfrak{a}^a \delta_b^\mathfrak{b}{A_a}^b = {\omega_\mathfrak{a}}^\mathfrak{b}$ and $\delta^\mathfrak{a}_{a}\theta^a = e^\mathfrak{a}\;$. Where the indices $\{\mathfrak{a,b,c,\ldots}\}$ belong to the tangent space of the effective deformed spacetime, $\omega$ is the spin connection 1-form and $e$ the vierbein 1-form. Thus, with the extra parametric identifications
\begin{equation} \label{eq:newton-cosmol}
\gamma^2=\frac{\kappa^2}{4\pi G}=\frac{4\Lambda^2}{3}\;,
\end{equation}
where $G$ is the Newtonian constant and $\Lambda^2$ is the renormalized cosmological constant\footnote{The renormalized cosmological constant has associated to the ans\"atz of the theory, what is, $\Lambda_{obs}^2=\Lambda_{qft}^2+\Lambda^2$.}. Hence, the action Eq.~\eqref{ym3} generates the following effective gravity action
\begin{equation}\label{ym-map-grav}
 S_{\mathrm{Grav}}=\frac{1}{16\pi G}\int \left\{\frac{3}{2\Lambda^2}{R_\mathfrak{a}}^\mathfrak{b}\star {R^\mathfrak{a}}_\mathfrak{b} -\frac{1}{2}\epsilon_\mathfrak{abcd}{R}^\mathfrak{ab} e^\mathfrak{c}e^\mathfrak{d} + T^\mathfrak{a}\star T_\mathfrak{a} + \frac{\Lambda^2}{12}\epsilon_\mathfrak{abcd}e^\mathfrak{a}e^\mathfrak{b}e^\mathfrak{c}e^\mathfrak{d}\right\}\;,
\end{equation}
where $R^\mathfrak{a}_{\phantom{a}\mathfrak{b}}=\mathrm{d}\omega^\mathfrak{a}_{\phantom{a}\mathfrak{b}}+\omega^\mathfrak{a}_{\phantom{a}\mathfrak{c}} \omega^\mathfrak{c}_{\phantom{c}\mathfrak{b}}$ and $T^\mathfrak{a}=\mathrm{d}e^\mathfrak{a}+\omega^\mathfrak{a}_{\phantom{a}\mathfrak{b}}e^\mathfrak{b}$ are, respectively, the curvature and torsion 2-forms. The symbol $\star$ stands for the Hodge dual operator in $\mathbb{M}^4$ (the deformed spacetime).

\newpage

\section{Running parameters and 1-loop estimates}\label{1LOOP}

Now\footnote{From now on, for the sake of simplicity, we use tensorial notation with Greek indices indicating spacetime coordinates and Latin indices to gauge group.}, we return to the original action \eqref{ym1} and we take only its quadratic part\footnote{See Appendix \ref{AP1} and references therein for details.}, considering also the Gribov-Zwanziger quadratic term\footnote{At this level, we are not considering the refined Gribov-Zwanziger action \cite{Dudal:2005na}.} \cite{Zwanziger1989},
\begin{eqnarray}\label{ym-quad}
&&S_{quad} = \int d^4x \left\{\frac{1}{4}\left(\partial_\mu Y_\nu^A - \partial_\nu Y_\mu^A \right)^2 + \frac{1}{2\alpha}(\partial_\mu Y_\mu^A)^2 +\bar{\varphi}_\mu^{AB}\partial^2 \varphi_\mu^{AB} + ~~~~~\right. \nonumber\\
&&\left. ~~~~~~~~~~~~~~~~~~~~-\lambda^2\kappa\left(f^{ABC} Y_\mu^A\varphi_\mu^{BC}+f^{ABC} Y_\mu^A\bar{\varphi}_\mu^{BC}  \right) -\lambda^4d\left[\frac{N(N-1)}{2}\right]\right\}\;,
\end{eqnarray}
where $\left(\varphi^{AB}_\mu,\overline{\varphi}^{AB}_\mu\right)$ is a pair of complex conjugate bosonic fields and $\lambda$ is, essentially, the Gribov parameter. Here we will use $N=5$ since we are building a Yang-Mills for the $SO(5)$ group. In a future we will employ this value, but, for now, we continue using $N$ in general. The paramter $\alpha$ is the gauge parameter associated with the gauge fixing. Inhere, the limit $\alpha\longrightarrow0$ must be employed in order to enforce the Landau gauge condition, \emph{i.e.} $\partial_\mu Y_\mu^A=0$. The choice of the gauge is, in principle, arbitrary as long as the gauge is renormalizable. However, most of the developments in the Gribov problem were made in the Landau gauge \cite{Gribov:1977wm,Zwanziger1989,Dudal:2005na,Dudal:2011gd}. Moreover, the Landau gauge is a very simple gauge to work with. Nevertheless, there are recent evidences that the Gribov parameter is a gauge invariant parameter \cite{Capri:2015pja,Capri:2015ixa,Capri:2016aqq,Pereira:2016fpn,Capri:2016aif}, a very welcome feature for the model.

\begin{figure}[h]
	\centering
		\includegraphics[width=0.4\textwidth]{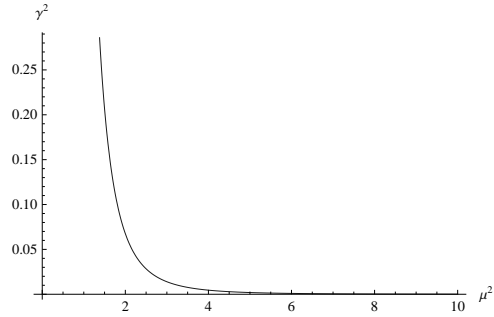}
		\caption{The running of the Gribov parameter as function of the energy scale squared. The energy squared $\mu^2$ is in units of $\overline{\Lambda}^2$ and the Gribov parameter is normalized in units of $(e^{5/6}/\sqrt{5})\overline{\Lambda}^2 $.}\label{fig1}
\end{figure}

At 1-loop, the effective action\footnote{Strictly speaking, Eq.~\eqref{ef-action} does not define the effective action since there are no source terms. In fact, $\Gamma^{(1)}$ is rather a field-independent shift of the actual $1$-loop effective action. This is sufficient for our purposes but does not constitute the full effective action. Nevertheless, since there is no formal term for it, we will call it simply by \textit{effective action} in this work. We expect no confusion by the reader.} is defined through
\begin{equation}\label{ef-action}
 e^{-\Gamma^{(1)}}=\int [D\Phi]e^{-S_{quad}}~,
\end{equation}
which, in $d$ dimensions, yields
\begin{equation}\label{gamma}
 \Gamma^{(1)}=-\lambda^4d\left[\frac{N(N-1)}{2}\right] + \frac{(d-1)}{2}\left[\frac{N(N-1)}{2}\right]\int \frac{d^d p}{(2\pi)^d}\left[ \ln\left( p^4+2N\kappa^2\lambda^4 \right) \right]~.
\end{equation}
To control the divergences of the quantum action we employ the $\overline{\mathrm{MS}}$ renormalization scheme to obtain
\begin{equation}\label{gamma-ren}
 \Gamma_r^{(1)}=-\lambda^4d\left[\frac{N(N-1)}{2}\right]-\frac{(d-1)}{32\pi^2}\left[\frac{N(N-1)}{2}\right](N\kappa^2\lambda^4)\left[\ln\left(\frac{2N\kappa^2\lambda^4}{\mu^4}\right)-\frac{8}{3} \right]~.
\end{equation}
where
\begin{equation}\label{lambda-gamma}
\gamma^4\equiv 2\kappa^2\lambda^4\;,
\end{equation}
is a more convenient mass parameter. At first sight, this choice is a mere algebraic \textit{ansatz} to simplify future computations with this parameter. In Appendix~\ref{AP2} we demonstrate why this choice is actually better than $\lambda$ for our purposes. Thus, for $d=4$, Eq.~\eqref{gamma-ren} turns to
\begin{equation}\label{gamma-ren2}
 \Gamma_r^{(1)}=-\frac{\gamma^4}{2k^2}4\left[\frac{N(N-1)}{2}\right]-\frac{3}{32\pi^2}\left[\frac{N(N-1)}{2}\right]\frac{N\gamma^4}{2}\left[\ln\left(\frac{N\gamma^4}{\mu^4}\right)-\frac{8}{3} \right]~.
\end{equation}
Following the Gribov-Zwanziger prescription \cite{Zwanziger:1989mf}, the Gribov parameter can be determined by minimizing the quantum action, i.e., $\partial\Gamma_r^{(1)}/\partial\gamma^2=0$. The result is
\begin{equation}\label{gamma-mu}
\frac{N\kappa^2}{16\pi^2}\left[\frac{5}{8}-\frac{3}{8}\ln\left(\frac{N\gamma^4}{\mu^4}\right)\right]=1~.
\end{equation}
Or, equivalently,
\begin{equation}
\gamma^2=\frac{e^\frac{5}{6}}{\sqrt{N}}\mu^2 e^{-\frac{4}{3}\left(\frac{16\pi^2}{N\kappa^2}\right)}\;.\label{gap2-gamma}
\end{equation}
Moreover, the $1$-loop coupling parameter is found to be \cite{Gross:1973id}
\begin{equation}
\frac{N\kappa^2}{16\pi^2}=\frac{1}{\frac{11}{3}\ln\frac{\mu^2}{\overline{\Lambda}^2}}\;,\label{kappa1}
\end{equation}
where $\overline{\Lambda}$ is the renormalization group cut-off. By inserting Eq.~\eqref{kappa1} into Eq.~\eqref{gap2-gamma} for $N=5$ we find
\begin{equation}
\gamma^2=\frac{e^\frac{5}{6}}{\sqrt{5}}\overline{\Lambda}^2\left(\frac{\mu^2}{\overline{\Lambda}^2}\right)^{-35/9}~.\label{gamma1}
\end{equation}
Thus, the higher the energy scale is, the smaller the Gribov parameter would be. This behaviour is plotted in Figure~\ref{fig1}.

\begin{figure}[htb]
	\centering
		\includegraphics[width=0.4\textwidth]{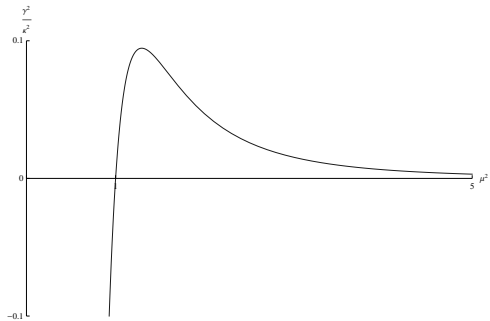}
		\caption{The running of the ratio $\gamma^2/\kappa^2$ as function of energy scale squared. The energy scale $\mu^2$ is in units of $\overline{\Lambda}^2$ and the ratio is in units of $\alpha\overline{\Lambda}^2$.}\label{fig2}
\end{figure}

As we have mentioned in Sec.~\ref{EFGRAV}, the ratio between the two quantum parameters, after we combine Eq.~\eqref{kappa1} and Eq.~\eqref{gamma1}, namely,
\begin{equation}
\frac{\gamma^2}{\kappa^2}=\alpha\overline{\Lambda}^2\left(\frac{\mu^2}{\overline{\Lambda}^2}\right)^{-35/9}\ln\left(\frac{\mu^2}{\overline{\Lambda}^2}\right)~,\label{ratio1}
\end{equation}
with $\alpha=55 e^{5/6}/\left(48\pi^2\sqrt{5}\right)$, is crucial for the present model. The behaviour of this ratio is illustrated in Figure~\ref{fig2}. It is clear that the expected behaviour $\gamma^2/\kappa^2\rightarrow0$ is attained at $\mu^2=\overline{\Lambda}^2$. 

\begin{figure}[h]
	\centering
		\includegraphics[width=0.4\textwidth]{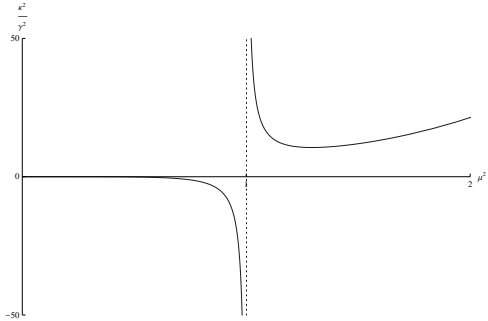}
		\caption{The ratio $\kappa^2/\gamma^2$ as a function of the energy scale squared. The energy scale is in units of $\overline{\Lambda}^2$ and the ratio is in units of $1/(\alpha\overline{\Lambda}^2)$.}\label{fig3}
\end{figure}

\newpage
The simple inversion of Eq.~\eqref{ratio1} gives
\begin{equation}
\frac{\kappa^2}{\gamma^2}=\frac{1}{\alpha\overline{\Lambda}^2}\left(\frac{\mu^2}{\overline{\Lambda}^2}\right)^{35/9}\left[\frac{1}{\ln\left(\frac{\mu^2}{\overline{\Lambda}^2}\right)}\right]\;,\label{ratio2}
\end{equation}
which shows the running behaviour of the ratio $\kappa^2/\gamma^2$ which is displayed in Figure~\ref{fig3}.
We remark that there is a discotinuity (Landau pole) at $\mu=\overline{\Lambda}$. We interpret this discontinuity as an indication of the transition between the quantum and classical regimes of the model. For $\mu<\overline{\Lambda}$ we expect a geometrical regime while for $\mu>\overline{\Lambda}$, the theory is at the quantum region. 

To estimate the Newtonian contant and the renormalization group cut-off we emphasize that we are not assigning any running behaviour to the Newtonian constant. Accordingly to Eq.~\eqref{eq:newton-cosmol}, our aim consists in identify the ratio $\gamma^2/\kappa^2$ with $G$ only after an energy scale is chosen. So the Newtonian constant is fixed as an effective quantity.

It is also important to realize that the deep infrared behaviour of Figure~\ref{fig1}, Figure~\ref{fig2} and Figure~\ref{fig3} do not reproduce the expected behaviour at zero momenta, as known by QCD lattice simulations \cite{Bloch:2003sk,Cucchieri:2006xi, Cucchieri:2009zt,Cucchieri:2012cb, Cucchieri:2014via,Cucchieri:2016jwg}. In the deep infrared regime the coupling parameter $\kappa$ goes to a finite value ,$~i.e.$, an infrared fixed point. However, this extreme behaviour is not relevant for the purpusoes of the present work.

\subsection{Numerical estimates at 1-loop}\label{NUMEST1}

We first follow the procedure performed in \cite{Assimos:2013eua} where the strategy was not to solve the gap equation by fixing $\overline{\Lambda}$ and $\mu$, which is the traditional way, but to fix the Newtonian constant and find if this is a consistent solution. Nevertheless, for the sake of consistency, we need a coupling constant as small as possible. Furthermore, we must have $\mu^2>\overline{\Lambda}^2$. Accordingly, we possess a certain range to work with, namely,
 \begin{eqnarray}\label{restrict-scales}
 0<\frac{N\kappa^2}{16\pi^2}<1~,&&\nonumber\\
 0<\ln\left(\frac{\mu^2}{\overline{\Lambda}^2}\right)<1~.&&
\end{eqnarray}
Let us take, for instance, $\mu^2=2\overline{\Lambda}^2$ in Eq.~\eqref{ratio2}, which provides $\ln(\mu^2/\overline{\Lambda}^2)=0.6931$, satisfying Ineq.~\eqref{restrict-scales}. One way to obtain the scale $\overline{\Lambda}$ is setting the Newtonian constant to its experimental value, i.e., $G =6.707\times10^{-33}~\mathrm{TeV}^{-2}$ in Eq.~\eqref{ratio2}, providing
\begin{equation}\label{lambda-bar1}
\overline{\Lambda}^2\approx 2.122\times10^{33}~\mathrm{TeV}^2\;~.
\end{equation}
This result allows us to estimate the renormalized cosmological constant. Combining Eq.~\eqref{eq:newton-cosmol}, Eq.~\eqref{gamma1} and Eq.~\eqref{lambda-bar1} we obtain
\begin{equation}\label{lambda1}
\Lambda^2\approx 1.106\times 10^{32}~\mathrm{TeV}^2\;~.
\end{equation}
We notice that the cut-off value \eqref{lambda-bar1} is just right above the Planck scale, given by $E_p^2=1.491\times 10^{32}\mathrm{TeV}^2$.

\subsubsection{Methods of enhancement at 1-loop}
\label{MethEnh1}

The main goal here is to calculate the best values for $N\kappa^2/16\pi^2$ and $\ln(\mu^2/\overline{\Lambda}^2)$ in accordance with Ineq.~\eqref{restrict-scales}. To handle this task we apply three methods, labelled by $M_1$, $M_2$ and $M_3$, as follows.

$M_1~:$\textbf{\emph{Taylor series method}}

Let us rewrite Eq.~\eqref{kappa1} as
\begin{equation}\label{taylor0}
\frac{1}{a}=\frac{11}{3}\ln b~,
\end{equation}
where
\begin{eqnarray}
 a&=&\frac{N\kappa^2}{16\pi^2}~, \nonumber\\
 b&=& \frac{\mu^2}{\overline{\Lambda}^2}~,
\end{eqnarray}
for merely simplification. Next, we expand the right hand side of Eq.~\eqref{taylor0} as a Taylor series at the critical point $\mu=\overline{\Lambda}$,$~i.e.~$, $b=1$ as follows
\begin{equation}\label{taylor1}
 \ln(b)= \sum_{n=1}^{\infty}\frac{1}{n}(-1)^{n-1}(b-1)^n~,
\end{equation}
with $0<\ln b<1$ as stated by Ineq.~\eqref{restrict-scales}. We investigate the series \eqref{taylor1} under two perspectives:
\begin{itemize}
\item Perspective (i): \textit{The endpoint extremum}

 The series expansion of $\ln(b)$ has radius of convergence equal to 1. Precisely, the alternating series test ensures that the series does not converge at $b=0$ and converges at $b=2$. Hence, the series is convergent for $0<b\leqslant 2$. Therefore, the endpoint extremum occurs at $b=2$ which happens also to be a global maximum. A curious fact about that series occurs when we truncate the series expansion at even $n^{th}$ order: any of these truncations have a maximum at $b=2$. Again such maximum is a global one and it happens at the endpoint. From Eq.~\eqref{taylor0}, it is clear that, by fixing a global maximum for $\ln(b)$, a minimum value for $N\kappa^2/16\pi^2$ is set. Thus, with $b=2$, we obtain $\ln(\mu^2/\overline{\Lambda}^2)=0.6931$ and $N\kappa^2/16\pi^2=0.3935$, which are both in accordance with the intervals described in Ineq.~\eqref{restrict-scales}.

\item Perspective (ii): \textit{A bound on the Taylor series for $\ln(b)$}

At this point, we are looking for a certain bound for the series expansion \eqref{taylor1}. Hence, and since $a<1$, we have
\begin{equation}\label{taylor3}
\ln(b)>3/11 \Longleftrightarrow \sum_{n=1}^{\infty}\frac{1}{n}(-1)^{n-1}(b-1)^n>3/11~.
\end{equation}
In this range we solve the Ineq.~\eqref{taylor3} for several $n$ values. We notice that the choices of $b$ values are restricted to $1.314<b<b_{sup}$ while $n$ is even and where $b_{sup}$ values decrease while even $n$ values increase. The $b_{sup}$ values are displayed in Table~\ref{tab1}.

\begin{table}[h!]
\centering\footnotesize
\begin{tabular}{c c }
 $n$ & $b_{sup}$ \\ 
\hline
 $2$ & $2.674$ \\ 
\hline
 $4$ & $2.476$ \\ 
\hline
 $8$ & $2.305$ \\ 
\hline 
 $10$ & $2.261$ \\ 
\hline
 $20$ & $2.158$ \\ 
\hline 
 $50$ & $2.079$ \\ 
\hline
 $100$ & $2.046$ \\ 
\hline 
 $1000$ & $2.007$ \\ 
\hline
 $5000$ & $2.002 $\\
\hline
 $10000$ & $2.000 $\\
\hline
\end{tabular}
\caption{The superior bound for $b$ range using only even values for $n$.} 
\label{tab1}
\end{table}
Still looking at Table~\ref{tab1}, for instance, $n=8\Rightarrow b_{sup}\approx 2.305$, $n=10\Rightarrow b_{sup}\approx 2.261$ and --- as actually we would expect --- $n\rightarrow\infty \Rightarrow b_{sup}\rightarrow 2.000$. However, if $n$ is odd we obtain for all intervals $b>1.313$, of course, and no upper bound. Hence, we have a confirmation of our first choice for $\ln(\mu^2/\overline{\Lambda}^2)$ given by $\mu^2=2\overline{\Lambda}^2$. 

Besides the best choice that we can perform, we still have freedom to choose any value consistent with the Ineq.~\eqref{restrict-scales}. If we pick, for instance, $a=0.4300$, we have $b=1.886\Rightarrow \ln(b)\equiv\ln(\mu^2\overline{\Lambda}^2)\approx 0.6342$, which provides, from Eq.~\eqref{ratio2} and Eq.~\eqref{eq:newton-cosmol},
\begin{equation}\label{lambda-bar2}
\overline{\Lambda}^2\approx 1.845\times10^{33} \mathrm{TeV}^2\;~
\end{equation}
and
\begin{equation}\label{lambda2}
\Lambda^2\approx 1.208\times 10^{32} \mathrm{TeV}^2\;~.
\end{equation}
These results can be interpreted as a numerical verification of the values Eq.~\eqref{lambda-bar1} and Eq.~\eqref{lambda1} due to the fact that their order of magnitude are maintained. In this sense we confirm the first insight presented in \cite{Assimos:2013eua}.
\end{itemize} 

$M_2~:$\textbf{\emph{Equilibrium value method}}

Now we use a simple method of enhancement which we called \emph{equilibrium value} between two functions at certain point. First, in order to simplify we take Eq.~\eqref{kappa1} as
\begin{equation}\label{fh}
 f(\kappa^2)h(\mu^2,\overline{\Lambda}^2)=\frac{3}{11}~,
\end{equation}
where
\begin{eqnarray}\label{nom-fh}
 f(\kappa^2)&=& \frac{N\kappa^2}{16\pi^2}~,\nonumber\\
 h(\mu^2,\overline{\Lambda}^2) &=& \ln\left(\frac{\mu^2}{\overline{\Lambda}^2}\right)~.
\end{eqnarray}
To obtain small values for $h(\mu^2,\overline{\Lambda}^2)$ and $f(\kappa^2)$, we made an equilibrium choice, $i.e.$, $h(\mu^2,\overline{\Lambda}^2)=f(\kappa^2)$. Consequently, it provides
\begin{equation}\label{hmin}
 h(\mu^2,\overline{\Lambda}^2)=\left(\frac{3}{11}\right)^\frac{1}{2}~~~\Rightarrow~~~\ln\left(\frac{\mu^2}{\overline{\Lambda}^2}\right)\approx 0.5222 ~~~\Rightarrow~~~\frac{\mu^2}{\overline{\Lambda}^2}\approx 1.686~.
\end{equation}
Thus, from Eq.~\eqref{ratio2} and Eq.~\eqref{eq:newton-cosmol}, we find
\begin{equation}\label{lambda-bar3}
\overline{\Lambda}^2\approx 1.449\times10^{33} \mathrm{TeV}^2\;~
\end{equation}
and
\begin{equation}\label{lambda3}
\Lambda^2\approx 1.468\times 10^{32} \mathrm{TeV}^2\;~.
\end{equation}
We conclude that \eqref{lambda-bar3} and \eqref{lambda3} do not show any significant improvement with respect to \eqref{lambda-bar1} and \eqref{lambda1}.

\newpage

$M_3~:$\textbf{\emph{ Method by geometrical series}}

Here we employ a geometrical series to treat the logarithm will be used. Due to Ineq.~\eqref{restrict-scales}, we can treat the logarithm in Eq.~\eqref{kappa1} as a geometrical series. First, we define
\begin{equation}\label{r-geo1}
r=1-\ln\frac{\mu^2}{\overline{\Lambda}^2}\;.
\end{equation}
Hence, we can use\footnote{We notice that $r<1$ due to Ineq.~\eqref{restrict-scales}.}
\begin{equation}\label{r-geo2}
\frac{1}{1-r}=\sum_{n=0}^{\infty} r^n
\end{equation}
in Eq.~\eqref{kappa1}, providing
\begin{equation}\label{kappa-geo}
\frac{N\kappa^2}{16\pi^2}=\frac{3}{11}\left(\frac{1}{1-r}\right)=\frac{3}{11}\sum_{n=0}^{\infty} r^n\;~.
\end{equation}
Second, we use Ineq.~\eqref{restrict-scales} and Eq.~\eqref{kappa-geo} to write
\begin{equation}\label{geo-ineq}
 \sum_{n=0}^{\infty} r^n < \frac{11}{3}~.
\end{equation}
Now, we test several truncations of expression Eq.~\eqref{geo-ineq} to deal with an $n^{th}$-degree polynomial inequality. Such procedure permits that we find $r\in (0,0.7273)$ as an optimum valid range. To clarify this point, for instance, we mount Table~\ref{tab2} displaying the evolution of this range, which directly determines the value of the logarithm.

\begin{table}[h!]
\centering\footnotesize
\begin{tabular}{c c }
 $n$ & $r_{sup}$ \\ 
\hline
 $5$ & $0.7974$ \\ 
\hline
 $8$ & $0.7470$ \\ 
\hline
 $10$ & $0.7367$ \\ 
\hline 
 $20$ & $0.7276$ \\ 
\hline
 $30$ & $0.7273$ \\ 
\hline 
 $40$ & $0.7273$ \\ 
\hline
 $100$ & $0.7273$ \\ 
\hline 
 $1000$ & $0.7273$ \\ 
\hline 
\end{tabular}
\caption{The superior bound $r_{sup}$ for the range of values for $r$ as a function of the of $n^{th}$-degree polynomial.}
\label{tab2}
\end{table}
We notice that $n>30$ does not bring any significant improvement for the superior bound of $r$. In this way, we choose $r\approx 0.7273$ as an optimal extreme valid value, which implies in $\ln(\mu^2/\overline{\Lambda}^2)\approx 0.2727$ and $N\kappa^2/16\pi^2\approx 0.3803$. With these values and using Eq.~\eqref{ratio2} and Eq.~\eqref{eq:newton-cosmol} we find the following results
\begin{equation}\label{lambda-bar4}
\overline{\Lambda}^2\approx 1.052\times10^{32} \mathrm{TeV}^2~,
\end{equation}
and
\begin{equation}\label{lambda4}
\Lambda^2\approx 2.810\times 10^{32} \mathrm{TeV}^2\;~.
\end{equation}
Then, $\overline{\Lambda}^2$ decreases in one order of magnitude when compared to \eqref{lambda-bar1}, \eqref{lambda-bar2} and \eqref{lambda-bar3}. It is straightforward to see how these values can be obtained directly from Eq.~\eqref{kappa-geo}. We stress out that the superior bound for $n<30$ leaves us with an invalid range for $N\kappa^2/16\pi^2$.

For the other extreme we choose $r=1.000\times 10^{-4}$, providing $\ln(\mu^2/\overline{\Lambda}^2)\approx 0.9999$ and $N\kappa^2/16\pi^2\approx 0.2728$. We use these values to find
\begin{equation}\label{lambda-bar5}
\overline{\Lambda}^2\approx 4.851\times10^{33} \mathrm{TeV}^2~
\end{equation}
and
\begin{equation}\label{lambda5}
\Lambda^2\approx 7.666\times 10^{31} \mathrm{TeV}^2\;~.
\end{equation}
In this case a better value for the renormalized cosmological constant is found. However, the renormalization group cut-off is the worst found until this point. To summarize all results that we found in each method we built Table~\ref{tab3}.

\begin{table}[h!]
\centering\footnotesize
\begin{tabular}{c c c c c c c }
& $I$ & $M_1$ & $M_2$ & $M_{3a}$ & $M_{3b}$ & $Pr$\\ 
\hline 
$\overline{\Lambda}^2(\mathrm{TeV}^2)$  & $2.122\times 10^{33}$ & $1.845\times 10^{33}$ & $1.449\times 10^{33}$ & $1.052\times 10^{32}$ & $4.851\times 10^{33}$ & $1.491\times 10^{32}$\\ 
\hline 
$\Lambda^2(\mathrm{TeV}^2)$  & $1.106\times 10^{32}$ & $1.208\times 10^{32}$ & $1.468\times 10^{32}$ & $2.810\times 10^{32}$ & $7.666\times 10^{31}$ & $3.710\times 10^{28}$\\ 
\hline 
\end{tabular}
\caption{The cut-off and renormalized cosmological constant obtained in each method. The column $I$ lists our initial estimates. The other columns $M_1$,~$M_2$,~$M_{3a}$ and $M_{3b}$ are related to the values obtained by Taylor series, equilibrium value and geometric series, respectively. The column $Pr$ exhibits the physical predictions, $i.e.$, the Planck energy squared and the absolute value predicted by the quantum field theory for the cosmological constant \cite{Weinberg:1988cp}.}
\label{tab3}
\end{table}

Comparing the numerical values for the cut-off and the renormalized cosmological constant which were obtained through the three methods $M_1$, $M_2$ and $M_3$ and listed in Table~\ref{tab3}, we observe that the orders of magnitude of those results are almost unchanged. An unique exception occurs to the cut-off in the column $M_{3b}$, which is caused by the extreme high value to the logarithm $\ln(\mu^2/\overline{\Lambda}^2)$.

\subsubsection{Fixing $\overline{\Lambda}$ as the Planck energy}\label{LBAR=PE}

We introduce here a different path to find values to Newtonian constant and the renormalized cosmological constant. In this manner we made all slightly different since we fit the cut-off $\overline{\Lambda}^2$ equals energy Planck,$~i.e~$, $\overline{\Lambda}^2=E_p^2=1.491\times 10^{16}~\mathrm{TeV}~$. Previously, in Sect.~\ref{NUMEST1} we found an optimum logarithm to fix the cut-off and the renormalized cosmological constant. With help  of fixed logarithms and Eq.~\eqref{ratio2} we compute the Newtonian constant $G_{p}$ for each method as displayed in Table~\ref{tab4}.
\begin{table}[htb]
\centering\footnotesize
\begin{tabular}{c c c c c c }
& $I$ & $M_1$ & $M_2$ & $M_{3a}$ & $M_{3b}$ \\ 
\hline 
$G_{p}(\mathrm{TeV}^{-2})$  & $ 9.551\times 10^{-32}$ & $8.301\times 10^{-32}$ & $6.521\times 10^{-32}$ & $5.254\times 10^{-32}$ & $2.183\times 10^{-31}$ \\ 
\hline 
 $\Lambda_{p}(\mathrm{TeV}^)$  & $ 7.766\times 10^{30}$ & $9.765\times 10^{30}$ & $1.510\times 10^{31}$ & $6.271\times 10^{31}$ & $2.355\times 10^{30}$ \\ 
\hline 
\end{tabular}
\caption{Newtonian constant and cosmological constant values based in the logarithm computed in each method in Sect.~\ref{NUMEST1} with the cut-off equals the Planck energy.}
\label{tab4}
\end{table}

We observe that all values for $G_p$ are in 1 order of magnitude above $G$. After confrontation with the values presented in Table~\ref{tab2} we notice a better estimate using method $M_{3b}$,$~i.e.~$, while we apply the logarithm obtained with the geometrical series for $\kappa^2$. The closer we can stay of $G=6.707\times 10^{-33}~\mathrm{TeV}^2$ happens when we apply the method $M_{3a}$. As consequence, the price to pay is a renormalized cosmological constant with a higher value than one encountered in the method $M_{3b}$. However, the order of magnitude is the same while we compare the values for $\Lambda^2$ found through the methods $M_{3a}$ and $M_{3b}$.

\newpage

\section{Numerical estimates at 2-loops}\label{2LOOP}

For 2-loops, an explicit analytical computation is a virtually impossible task. To work out this equation, sophisticated algebraic programs were built. For instance, \textbf{FORM} and \textbf{QGraph} programs are frequently used as well as the recent developments about such computational packages \cite{Nogueira:1991ex,Ford:2009ar,Kuipers:2012rf,Kuipers:2013pba,Ueda:2014sya,Steinhauser:2015wqa}. In this section we borrow the main results of 2-loops computations from \cite{Ford:2009ar}.

\subsection{2-loops $\beta$-function}

\begin{figure}[htb]
	\centering
		\includegraphics[width=0.7\textwidth]{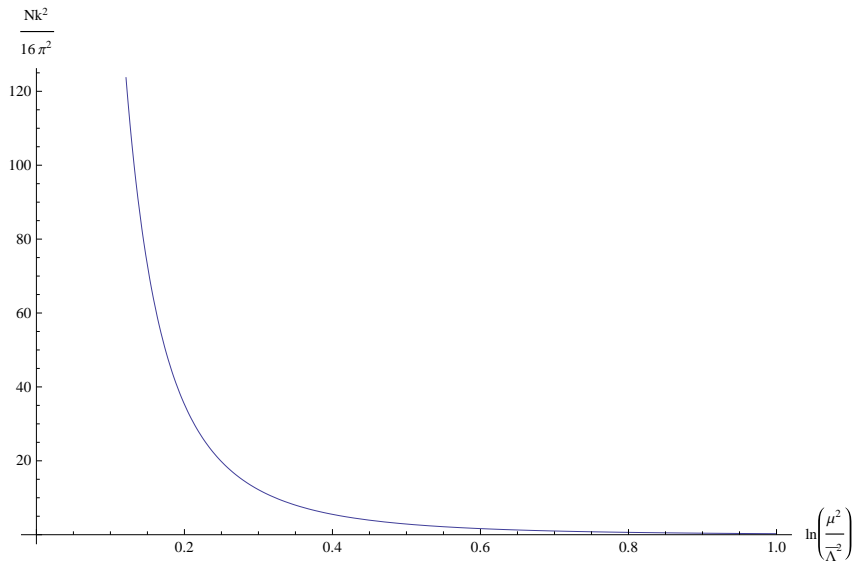}
		\caption{The behaviour of the coupling parameter related to the energy scale.}\label{fig7}
\end{figure}

First, recalling that the $\beta$-function at 2-loops \cite{Ford:2009ar,Collins:1984xc} is given by
\begin{equation}\label{beta-2loop}
\beta(\kappa^2) = - \frac{11N}{3}\left(\frac{\kappa^2}{16\pi^2}\right)^2 - \frac{34}{3} N^2 \left(\frac{\kappa^2}{16\pi^2}\right)^3~,
\end{equation}
the 2-loops running coupling constant is
\begin{equation}\label{kappa3}
\frac{N\kappa^2}{16\pi^2}=\frac{1}{\frac{11}{3}\ln\left(\frac{\mu^2}{\overline{\Lambda}^2}\right)}
 - \frac{102}{121}\left\{\frac{\ln\left[\ln\left(\frac{\mu^2}{\overline{\Lambda}^2}\right)\right]}{\left[\ln\left(\frac{\mu^2}{\overline{\Lambda}^2}\right)\right]^2}\right\} ~,
\end{equation}
where $\overline{\Lambda}$ is the cut-off of the energy scale. The evolution of the coupling $\kappa$, related to energy scale $\mu$, is displayed in Fig.~\eqref{fig7}.

\newpage

\subsection{2-loops gap equation}

From \cite{Ford:2009ar}, the main result is the 2-loops Gribov gap equation in the $\overline{\mathrm{MS}}$ with massive quarks.
 Here, we are dealing with a theory without fermions. Hence, from \cite{Ford:2009ar}, the 2-loops gap equation reduces to the simpler form
\begin{eqnarray} 
1 &=& \left(\frac{N\kappa^2}{16\pi^2}\right)\left[ \frac{5}{8} - \frac{3}{8} \ln\left( \frac{N\gamma^4}{\mu^4} \right) \right] +\nonumber \\ 
&& + \left(\frac{N\kappa^2}{16\pi^2}\right)^2\left\{ \frac{3893}{1536} + \frac{825}{4096} \sqrt{3}\pi^2
+ \frac{29}{768}\pi^2 - \frac{65}{48} \ln\left(\frac{N\gamma^4}{\mu^4} \right)
+ \frac{35}{128}\ln^2\left(\frac{N\gamma^4}{\mu^4} \right)+ \right. \nonumber \\
&& \left. \left. + \frac{137}{2048}\sqrt{5}\pi^2 - \frac{1317}{4096}\pi^2 \right\} \right.~.
\label{gap2loops}
\end{eqnarray} 
First of all, we compute the system formed by Eq.~\eqref{kappa3} and Eq.~\eqref{gap2loops} to analyze the behavior of the Gribov parameter related to energy scale. Such procedure gives to us the following two functions, where we have now the mass parameter related to logarithm.
\begin{eqnarray}\label{gamma-mu-2loop}
\gamma_m^2&=&\frac{1}{\sqrt{5}}\mu^2\left[h(\mu)\right]^{-\mathcal{H}(\mu)} e^{\mathcal{W}_m(\mu)}~,\nonumber\\
\gamma_p^2&=&\frac{1}{\sqrt{5}}\mu^2\left[h(\mu)\right]^{-\mathcal{H}(\mu)} e^{\mathcal{W}_p(\mu)}~,
\end{eqnarray}
where
\begin{eqnarray}\label{functions-mu}
&&h(\mu)=\ln\left(\frac{\mu^2}{\overline{\Lambda}^2}\right)~,\nonumber\\
&&\mathcal{H}(\mu)=\frac{1496}{105}\frac{\mathcal{P(\mu)}}{\mathcal{Q(\mu)}}~,\nonumber\\
&&\mathcal{P}(\mu)=h(\mu)\left[33h(\mu)+65\right]~,\nonumber\\
&&\mathcal{Q}(\mu)=\left\{11h(\mu)-34\ln\left[h(\mu)\right]\right\}^2~,\nonumber\\
&&\mathcal{W}_m (\mu)=\frac{1}{1680\mathcal{Q}(\mu)}\left[\mathcal{S}(\mu)-\sqrt{2}\mathcal{T}(\mu)\right]~,\nonumber\\
&&\mathcal{W}_p (\mu)=\frac{1}{1680\mathcal{Q}(\mu)}\left[\mathcal{S}(\mu)+\sqrt{2}\mathcal{T}(\mu)\right]~,\nonumber\\
&&\mathcal{S}(\mu)=a_1h^3(\mu)+a_2h^2(\mu)+a_3\ln^2\left[h(\mu)\right] ~,\nonumber\\
&&\mathcal{T}(\mu)=\sqrt{\mathcal{Q}(\mu)\left\{
b_1h^4(\mu)+ b_2h^3(\mu)-b_3h^2(\mu)-b_4\ln^2\left[h(\mu)\right] + b_5\ln\left[h(\mu)\right]h(\mu)-b_6\ln\left[h(\mu)\right]h^2(\mu)
\right\}}~,\nonumber\\
\end{eqnarray}
where $a_1=255,552$, $a_2=251,680$, $a_3=2,404,480$, $b_1=2,368,796,672$, $b_2=173,775,360$, $b_3=605d_0$, $b_4=5780d_0$, $b_5=3740d_0$, $b_6=537,123,840$, $d_0=221,384+b_0$ and $b_0= 21\left(-3,487+2,475\sqrt{3}+822\sqrt{5}\right)\pi ^2$.

\begin{figure}[htb]
	\centering
		\includegraphics[width=0.5\textwidth]{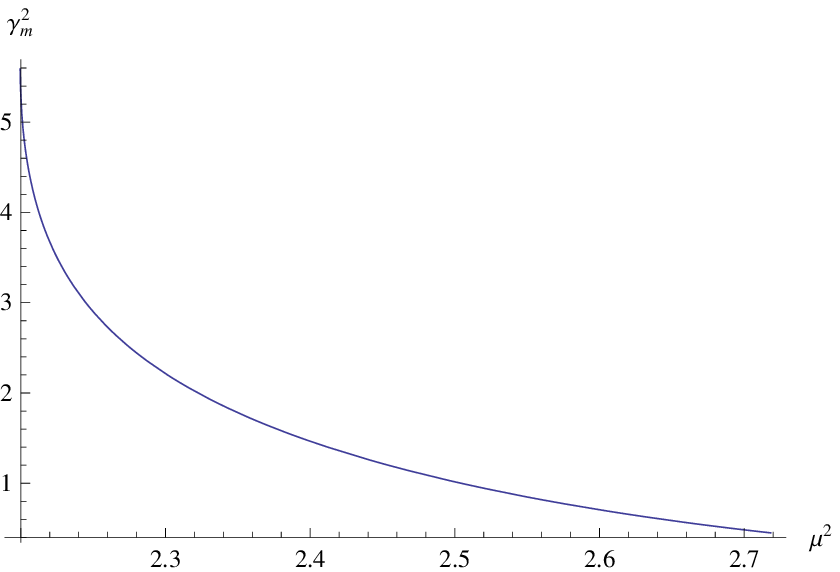}
		\caption{Gribov parameter $\gamma_m^2$ as a function of the energy scale $\mu^2$. Both $\gamma_m^2$ and $\mu^2$ are in units of $\overline{\Lambda}^2$.}\label{fig8}
\end{figure}

\begin{figure}[htb]
	\centering
		\includegraphics[width=0.5\textwidth]{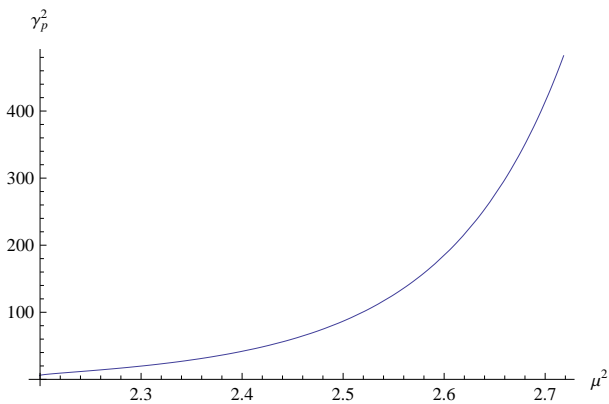}
		\caption{Gribov parameter $\gamma_p^2$ as a function of the energy scale $\mu^2$. Both $\gamma_p^2$ and $\mu^2$ are in units of $\overline{\Lambda}^2$.}\label{fig9}
\end{figure}
The behavior of $\gamma_m^2$ and $\gamma_p^2$ in Eq.~\eqref{gamma-mu-2loop} can be clearly seen in Figure \eqref{fig8} and Figure \eqref{fig9}, respectively. The behavior of  $\gamma_p^2$ and $\gamma_m^2$ indicates uniquely $\gamma_m^2$ as the one that has the expected typical running behavior of a mass parameter in the Gribov-Zwanziger scenario. Thus, we necessarily keep $\gamma_m^2$ for the next computations.

\subsection{Methods of enhancement at 2-loops}
\label{MethEnh2}

Following similar steps that we have made in Sec.~\ref{MethEnh1}, we are looking for the best logarithm for the sake of better estimates of the renormalized cosmological constant $\Lambda^2$ and the energy cut-off $\tilde{\Lambda}^2$. Before we advance in applying these methods, we refer to the definitions \eqref{nom-fh}. Moreover, we skip the initial choice, as made at 1-loop,~$\mu^2/\overline{\Lambda}^2=2$ because it provides $N\kappa^2/16\pi^2=1.036$, which is outside the acceptable range for the coupling parameter, see Ineq.~\eqref{restrict-scales}. Hence, we proceed with the methods of enhancement as we have done in Sect.~\ref{MethEnh1}.

$M_1$:~\textbf{\textit{Taylor series method}}

We are looking for small logarithms through a Taylor expansion of Eq.~\eqref{kappa3}, which can be written as
\begin{equation}\label{kappa-taylor}
 f(\kappa)=\frac{3}{11} -\frac{135}{121}(h(\mu)-1) + \frac{288}{121}(h(\mu)-1)^2 - \frac{475}{121}(h(\mu)-1)^3 + \frac{125}{22}(h(\mu)-1)^4 - \frac{4602}{605}(h(\mu)-1)^5 + \mathcal{O}((h(\mu)-1)^6)\;.
\end{equation}
where, for simplicity, the expansion above is displayed up to fifth order. However, we must keep in mind that we can truncate such expansion at any arbitrary order. If we truncate the above expansion of $f(\kappa)$ at fourth order, and consider $0<f(\kappa)<1$, then we find $h(\mu)>0.6938$. Therefore, we obtain $0.6938 < h(\mu)<1.000$. All truncations beyond the fourth order do not imply in significant improvements in the inferior limit $h_{inf}$, in the interval $h_{inf}<h(\mu)<1$, of the intervals for $h(\mu)$, since each order of truncation modifies such limit (See Table~\ref{tab5}).

\begin{table}[h!]
\centering\footnotesize
\begin{tabular}{c c }
 $n_t$ & $h_{inf}$ \\ 
\hline
 $2$ & $0.6340$ \\
 \hline
 $3$ & $0.6806$ \\
\hline
 $4$ & $0.6938$ \\ 
\hline
 $5$ & $0.6983$ \\
 \hline
 $6$ & $0.6998$ \\
 \hline
 $7$ & $0.7004$ \\
\hline
 $8$ & $0.7006$ \\ 
\hline 
 $10$ & $0.7007$ \\ 
\hline
 $50$ & $0.7007$ \\ 
\hline 
 $100$ & $0.7007$ \\ 
\hline
 $500$ & $0.7007$ \\ 
\hline 
 $1000$ & $0.7007$ \\ 
\hline 
\end{tabular}
\caption{The inferior limit $h_{inf}$ for the range of values for $h(\mu)$ accordingly to $n_t$, which is the order of truncation of the expansion for $f(\kappa)$.}
\label{tab5}
\end{table}

Since we are dealing with a perturbation expansion, we need to get to small values for $h(\mu)$. Nevertheless, because of the second term in Eq.~\eqref{kappa3}, which is resulting from the contribution at 2-loops in the computation, a choice for the logarithm close to any $h_{inf}$ results in high numerical values of the renormalized cosmological constant $\Lambda^2$. Hence, the best choice for the logarithm is $\ln(\mu^2/\overline{\Lambda}^2)=0.9999$. Employing this logarithm value and combining Eq.~\eqref{kappa3}, Eq.~\eqref{gap2loops} and Eq.~\eqref{eq:newton-cosmol}, we obtain
\begin{equation}\label{lambdabar5}
\overline{\Lambda}^2\approx 2.269\times10^{32} \mathrm{TeV}^2
\end{equation}
and
\begin{equation}\label{ccosmo2loops}
\Lambda^2\approx 7.665\times 10^{31}\mathrm{TeV}^2~.
\end{equation}
The value \eqref{lambdabar5} is very close to the order of magnitude of the Planck energy $E_p^2$. The result \eqref{lambda-bar5} is almost better than we found in Sec.~\ref{MethEnh1} at 1-loop approximation. The result \eqref{ccosmo2loops} certifies the 1-loop result for $\Lambda^2$.

$M_2$: \textbf{\textit{Equilibrium value method}}. If we apply $f(\kappa)=h(\mu)$ in Eq.~\eqref{kappa3}, we obtain the result $h(\mu)=0.7599$. However, this value, even though it obeys Ineq.~\eqref{restrict-scales}, it does not provide a real value for the Gribov parameter $\gamma^2_m$, according to Eq.~\eqref{gap2loops}.

$M_3$:~\textbf{\textit{Method by geometric series}}

In this case, we will treat the logarithm as a geometric series. For such aim, we employ once again Eqs.~\eqref{r-geo1} and \eqref{r-geo2} by treating $r$ as the ratio of a geometric series. In this way, we combine Eq.~\eqref{r-geo1}, Eq.~\eqref{r-geo2} and Eq.~\eqref{kappa3} to find
\begin{equation}
f(\kappa)=\frac{3}{11}\sum^{n_t}_{n=0}r^n+\frac{102}{121} \left(\sum^{n_t}_{n=0}r^n\right)^2 \ln\left(\sum^{n_t}_{n=0}r^n\right)~,
\end{equation}
where $n_t$ is the truncation order. Now, we must solve the inequality $0<f(\kappa)<1$ to find the range for valid logarithms. Each $n_t$ results in an inequality in the form $0<r<r_{sup}$. To clarify this point, the superior limit $r_{sup}$ as a function of $n_t$ is displayed at Table~\ref{tab7}.
\begin{table}[h!]
\centering\footnotesize
\begin{tabular}{c c }
 $n_t$ & $r_{sup}$ \\ 
 \hline
 $3$ & $0.3054$ \\
\hline
 $4$ & $0.3011$ \\ 
\hline
 $5$ & $0.2998$ \\
\hline
 $6$ & $0.2995$ \\
\hline
 $7$ & $0.2994$ \\
\hline
 $8$ & $0.2994$ \\
\hline
 $9$ & $0.2993$ \\
\hline
 $10$ & $0.2993 $ \\ 
\hline 
\end{tabular}
\caption{The superior limit $r_{rup}$ accordingly to $n_t$, which is the order of truncation of the expansion for $f(\kappa)$.}
\label{tab7}
\end{table}

The truncation at fourth order works just as good as the 1-loop case, because the values for the superior limit for $r$ do not reveal any significant changes. At first sight, we could work with the interval $0<r<0.2993$, however there is another constraint due to the Gribov parameter function $\gamma_m^2$, according to Eq.~\eqref{gamma-mu-2loop} and Eq.~\eqref{functions-mu}. The square root in $\gamma_m^2$ only has a real solution if $0.7882<h(\mu)<1$, or equivalently, $0<r<0.2117$. This interval is actually more restrictive than the $1$-loop treatment. Therefore, we are satisfied with the value $r=0.2117$. Hence, using $h(\mu)=0.7883$ in the equation system formed by Eq.~\eqref{kappa3}, Eq.~\eqref{gap2loops} and Eq.~\eqref{eq:newton-cosmol}, we get
\begin{equation}\label{lambdabarGS}
\overline{\Lambda}^2\approx 4.593\times10^{31} \mathrm{TeV}^2
\end{equation}
and
\begin{equation}\label{ccosmo2loopsGS}
\Lambda^2\approx 1.879\times 10^{32}\mathrm{TeV}^2~.
\end{equation}

On the other hand, for the other extreme, we choose $r=1.000\times 10^{-4}$ and the same equation system (Eqs.~\eqref{kappa3}, \eqref{gap2loops} and \eqref{eq:newton-cosmol}), we find
\begin{equation}\label{lambdabarGS}
\overline{\Lambda}^2\approx 2.269\times10^{32} \mathrm{TeV}^2
\end{equation}
and
\begin{equation}\label{ccosmo2loopsGS}
\Lambda^2\approx 7.665\times 10^{31}\mathrm{TeV}^2~.
\end{equation}
The values \eqref{lambdabarGS} and \eqref{ccosmo2loopsGS} match with the first ones that we found using $M_1$.

\subsection{The logarithmic elimination option}

A straight and simple manner to simplify the gap equation Eq.~\eqref{gap2loops} consists in setting up all logarithms to zero. For this purpose can set
\begin{equation}
\frac{\gamma^2}{\mu^2}=\frac{1}{\sqrt{5}}\;.
\end{equation}
Thus, Eq.~\eqref{gap2loops} can easily be solved, providing $N\kappa^2/16\pi^2\approx 0.4013$ and $\ln(\mu^2/\overline{\Lambda}^2)\approx 0.9067$. These results leave us to the following 2-loops values for the energy cut-off and the renormalized cosmological constant,
\begin{equation}\label{lambda-bar6}
\overline{\Lambda}^2\approx 1.066\times 10^{30} \mathrm{TeV}^2
\end{equation}
and
\begin{equation}\label{lambda6}
\Lambda^2\approx 3.589\times 10^{31} \mathrm{TeV}^2~.
\end{equation}

We remark that, at 1-loop, this strategy is not permitted because the value for the coupling parameter is found to be higher than 1, namely $N\kappa^2/16\pi^2=1.600$.

\section{Conclusions}\label{FINAL}

The Gribov mass parameter in our theory is the central point in the development of the induced gravity discussed in \cite{Sobreiro:2011hb}. It is responsable to address the deformation of the Yang-Mills theory in the infrared regime to a geometrical theory of gravity. In this model, the Gribov parameter and the Yang-Mills coupling constant combine in order to provide the value of Newtonian constant. In the present work we have developed 1 and 2-loops estimates to accomodate reliable values for the prediction of Newtonian constant. Moreover, the renormalization group scale could also determined. Furthermore, a discussion about the cosmological constant was performed.

Our results show that the experimental value of the Newtonian constant is a solution of the theory and that the renormalization group scale always lie around Planck scale, a good feature for a quantum gravity model candidate. We have also improved the estimates in order to attain the best values for these constantes, at 1 and 2-loops. 

Concerning the cosmological constant problem, our model provides an inherent gravitational cosmological constant. Following \cite{Shapiro:2006qx,Shapiro:2009dh}, it would be nice that this constant, added to the QFT prediction for the Standard Model vacuum \cite{Weinberg:1988cp} would compensate each other and provide a very small value for an observational cosmological constant. However, although very high, our cosmological constant differs from the QFT prediction by a few orders in magnitude.

Of course, much has to be investigated. For instance, since the emergence of gravity relies on the Gribov parameter and the soft BRST breaking \cite{Baulieu:2008fy,Baulieu:2009xr,Dudal:2012sb}, the Gribov-Zwanziger refinement \cite{Dudal:2011gd,Capri:2015nzw,Capri:2016aqq} should also be considered with all the extra mass parameters. These extra masses should also refine the values we have computed in the present work.  One interesting feature that should be mentioned of the theory is about the renormalized cosmological constant: first of all, at the quantum level, it is related to the Gribov parameter, so it is a running mass parameter; second, at classical level, its value does not run anymore. Its huge fixed value takes place at the classical sector of the theory. This is very important because it suppresses the quadratic curvature term in Eq.~\eqref{ym-map-grav}, at classical level, ensuring the general relativity limit of the theory, at least for a torsionless regime.

\appendix

\section{The Gribov-Zwanziger action}\label{AP1}

We will present here a brief description on the Gribov-Zwanziger scenario. For the details of technicalities and fundamental concepts we refer to \cite{Gribov:1977wm,Singer:1978dk,Zwanziger1989,Zwanziger:1989mf,Sobreiro:2005ec,Dudal:2005na,Sorella:2009vt,Dudal:2009bf}.

Quantization of Yang-Mills theories is a hard work. Initially, the procedure established by Faddeev and Popov \cite{Faddeev:1967fc} was well-succeeded in the perturbative regime during the process of quantizing the gauge fields. However, the Faddeev-Popov method is not accurate at low energies, where the system becomes highly non-perturbative. In essence, a gauge symmetry survives and is manifest at the infrared region. This is the so called Gribov ambiguities problem. The way to treat such ambiguities, as proposed by Gribov, is to look for a region in the gauge field space without ambiguities and truncate the Faddeev-Popov path integral to such region. Such a region is called \textit{fundamental modular region}. However the implementation of such region in the Faddeev-Popov path integral is a highly nontrivial problem with no solution so far. Nevertheless, the problem can be partially solved by restricting the path integral to the so called Gribov region, which is well defined for only a few gauges such as the Landau gauge. At the Landau gauge, the Gribov region can be defined as
\begin{equation}
\Omega=\{Y^A_\mu,\partial_\mu Y^A_\mu=0,M^{AB}>0~, \}
\end{equation}
with $D^{AB}_\mu=\delta^{AB}\partial_\mu-gf^{ABC} Y^B Y^C$ and $M^{AB}=-\partial_\mu D_\mu^{AB}$ is the Faddeev-Popov operator. Following for instance \cite{Zwanziger:1989mf,Sobreiro:2005ec,Sorella:2009vt,Dudal:2009bf}, the improved gauge fixed Faddeev-Popov action is given by the Gribov-Zwanziger action
\begin{equation}
S_{GZ}=S_{YM}+S_{gf}+\int d^4x \gamma^4 g^2f^{ABC}Y^B_\mu\overline{M}^{AB}f^{DEC}Y^E_\mu~+\int d^4x 4\gamma^4 (N^2-1)\;,
\end{equation}
where, $M^{AB}(x)\overline{M}^{BC}(x,y)=\delta^4(x-y)\delta^{AC}$. The non-local term is known as the horizon function and, together with the gap equation,
\begin{equation}\label{AAA}
\frac{\delta\Gamma}{\delta\gamma^2}=0\;,
\end{equation}
ensures that the path integral is inside the Gribov region. In \eqref{AAA}, $\Gamma$ is the quantum action, determined by
\begin{equation}
e^{-\Gamma}=\int [d\Psi]e^{-S_{GZ}}~.
\end{equation}

Remarkably, the non-local term can be written in local form with the help of auxiliary fields by means of
\begin{equation}
e^{-S_h}=\int \left[d\Phi\right]e^{-S_{loc}}~,
\end{equation}
with $\left[d\Phi\right]\equiv[d\varphi][d\overline{\varphi}][d\omega][d\overline{\omega}]$ and
\begin{equation}
S_{loc}=\int d^4x\left[-\overline{\varphi}_\mu^{AC}M^{AB}\varphi_\mu^{BC}+\overline{\omega}_\mu^{ac}M^{ab}\overline{\omega}_\mu^{bc}-\gamma^2 gf^{ABC}Y^A_\mu\left(\varphi^{BC}_\mu+\varphi^{BC}_\mu\right)\right]~,
\end{equation}
where the conjugate complex pair $(\overline{\varphi}_\mu^{AC},\varphi_\mu^{BC}$) are bosonic fields and  $(\overline{\omega}_\mu^{AC}\omega_\mu^{AC})$ are fermionic fields. Hence, the local version of the Gribov-Zwanziger path integral is
\begin{equation}\label{AAA1}
\mathcal{Z}=\int [d\Psi] e^{-S_{YM}-S_{gf}-S_{loc}-\int d^4x 4\gamma^4 (N^2-1)~.}~,
\end{equation}
with $\left[d\Psi\right]\equiv [dY][d\varphi][d\overline{\varphi}][d\omega][d\overline{\omega}][db][dc][d\overline{c}]$

The quadratic action \eqref{ym-quad}, as mentioned in Sect.~\ref{1LOOP}, is obtained from the free part of \eqref{AAA1}. We can straightforwardly observe that the fermionic fields do not contribute in the quadratic action. It is worth mentioning that many contributions from condensates \cite{Dudal:2011gd,Capri:2013naa} also appear in a refined version of \eqref{AAA1}, however, they are not relevant for the purposes of this work.

\section{The choice of the mass parameter}\label{AP2}

\begin{figure}[htb]
	\centering
		\includegraphics[width=0.6\textwidth]{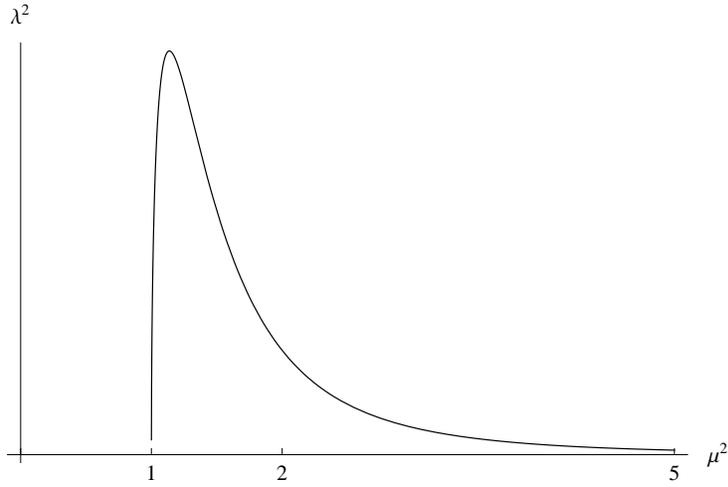}
		\caption{Gribov parameter as function of energy scale. The energy is in units of $\overline{\Lambda}$ and the $\lambda^2$ parameter in units of $\xi\overline{\Lambda}^2$.}\label{fig4}
\end{figure}

Although we are satisfied with the results obtained in this work, one can argue if another parameter could describe the Newtonian constant and the cosmological constant in a better way. Since we are not considering any condensation effect, the other possibility would be $\lambda$ rather than $\gamma$. Let us reconsider \eqref{lambda-gamma} and rewrite the gap equation taking into account $\lambda$ instead of $\gamma$,
\begin{equation}
\lambda^2=e^\frac{5}{6}\frac{\mu^2}{(2N\kappa^2)^{1/2}}e^{-\frac{4}{3}\left(\frac{16\pi^2}{N\kappa^2}\right)}\;.\label{gap2-lambda}
\end{equation}
Manipulating Eq.~\eqref{kappa1} and Eq.~\eqref{gap2-lambda} we get
\begin{equation}
\lambda^2=\xi\overline{\Lambda}^2\left(\frac{\mu^2}{\overline{\Lambda}^2}\right)^{-35/9}\ln^{1/2}\left(\frac{\mu^2}{\overline{\Lambda}^2}\right)\;,\label{lambda-g}
\end{equation}
with $\xi=e^{5/6}\sqrt{11/96\pi^2}$. 
And Eq.~\eqref{lambda-g} points again that the smaller the Gribov parameter, the higher the energy scale. We can note in Figure~\ref{fig4} the behaviour of the mass parameter $\lambda^2$ when the energy scale rises.
Here, we obtain $\lambda^2=0$ when $\mu=\overline{\Lambda}$. There is another troublesome here because the mass parameter $\lambda^2$ shows itself with the local maximum at $\mu=e^{9/140}\overline{\Lambda}$ which indicates \textemdash before this point \textemdash a decreasing of the mass parameter while the energy scale decreases too. It is completely antagonic to its physical behaviour at low energy regime where we expect a monotonous rising of the mass parameter while the energy decreases.

\begin{figure}[h!]
	\centering
		\includegraphics[width=0.6\textwidth]{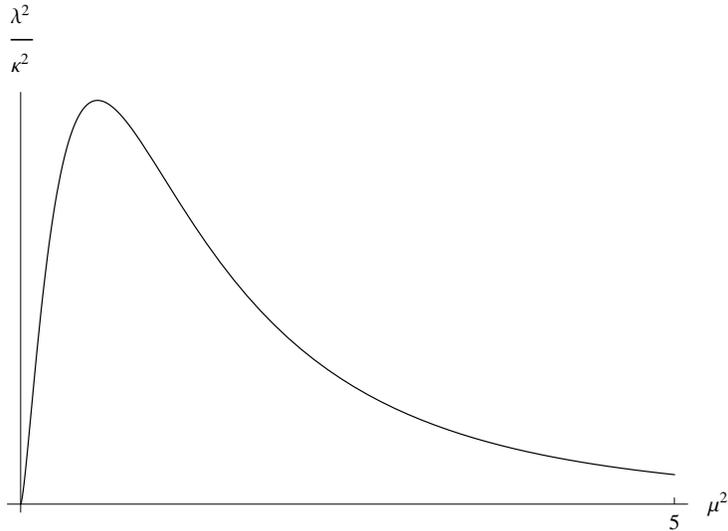}
		\caption{The ratio $\lambda^2/\kappa^2\equiv (4\pi G)^{-1}$ as function of energy scale $\mu$. The energy scale is in units of $\overline{\Lambda}^2$ and the $\lambda^2/\kappa^2$ is in units of $\rho\overline{\Lambda}^2$.}\label{fig5}
\end{figure}

From Eq.~\eqref{lambda-gamma}, Eq.~\eqref{kappa1} and Eq.~\eqref{gamma1} we get
\begin{equation}
\frac{\lambda^2}{\kappa^2}=\rho\overline{\Lambda}^2\left(\frac{\mu^2}{\overline{\Lambda}^2}\right)^{-35/9}\ln^{\frac{3}{2}}\left(\frac{\mu^2}{\overline{\Lambda}^2}\right)~,\label{ratio3}
\end{equation}
with $\rho= (55/192\pi^3)e^{5/6}\sqrt{11/6}$.
The behaviour of the ratio given by Eq.~\eqref{ratio3} can be ilustrated by the Figure \eqref{fig5}.
Figure~\eqref{fig5} clearly shows a nonexistence of a transition because an In\"on\"u-Wigner contraction can not happen. The vanishing limit of $\lambda^2/\kappa^2$ only occurs at the origin, hence, no geometric sector would appear. Then, we made the right choice to employ the Gribov parameter $\gamma$ in order to fit with our theory.
\newpage

\section*{Acknowledgements}

The Coordena\c c\~ao de Aperfei\c coamento de Pessoal de N\'ivel Superior (CAPES) and the Pr\'o-Reitoria de Pesquisa, P\'os-Gradua\c c\~ao e Inova\c c\~ao (PROPPI-UFF) are acknowledge.


\begin{thebibliography}{99}

\bibitem{Ashtekar:2004eh} 
  A.~Ashtekar and J.~Lewandowski,
  ``Background independent quantum gravity: A Status report'',
  Class.\ Quant.\ Grav.\  {\bf 21}, R53 (2004)
  doi:10.1088/0264-9381/21/15/R01
  [gr-qc/0404018].
  
\bibitem{Rovelli:2014ssa} 
  C.~Rovelli and F.~Vidotto,
  ``Covariant Loop Quantum Gravity : An Elementary Introduction to Quantum Gravity and Spinfoam Theory'',

\bibitem{Asorey:1996hz} 
  M.~Asorey, J.~L.~Lopez and I.~L.~Shapiro,
  ``Some remarks on high derivative quantum gravity'',
  Int.\ J.\ Mod.\ Phys.\ A {\bf 12}, 5711 (1997)
  doi:10.1142/S0217751X97002991
  [hep-th/9610006].
  
\bibitem{Buchbinder:1992rb} 
  I.~L.~Buchbinder, S.~D.~Odintsov and I.~L.~Shapiro,
  ``Effective action in quantum gravity'',
  Bristol, UK: IOP (1992) 413 p

\bibitem{Henson:2006kf} 
  J.~Henson,
  ``The Causal set approach to quantum gravity'',
  In *Oriti, D. (ed.): Approaches to quantum gravity* 393-413
  [gr-qc/0601121].

\bibitem{Ambjorn:2012jv} 
  J.~Ambjorn, A.~Goerlich, J.~Jurkiewicz and R.~Loll,
  ``Nonperturbative Quantum Gravity'',
  Phys.\ Rept.\  {\bf 519}, 127 (2012)
  [arXiv:1203.3591 [hep-th]].

\bibitem{Witten:1998qj} 
  E.~Witten,
  ``Anti-de Sitter space and holography'',
  Adv.\ Theor.\ Math.\ Phys.\  {\bf 2}, 253 (1998)
  [hep-th/9802150].

\bibitem{Seiberg:1999vs} 
  N.~Seiberg and E.~Witten,
  ``String theory and noncommutative geometry'',
  JHEP {\bf 9909}, 032 (1999)
  [hep-th/9908142].

\bibitem{Hawking:1979ig} 
  S.~W.~Hawking and W.~Israel,
  ``General Relativity : An Einstein Centenary Survey,''

\bibitem{Reuter:2012id} 
  M.~Reuter and F.~Saueressig,
  ``Quantum Einstein Gravity'',
  New J.\ Phys.\  {\bf 14}, 055022 (2012)
  [arXiv:1202.2274 [hep-th]].

\bibitem{Barcelo:2001tb} 
  C.~Barcelo, M.~Visser and S.~Liberati,
  ``Einstein gravity as an emergent phenomenon?,
  Int.\ J.\ Mod.\ Phys.\ D {\bf 10}, 799 (2001)
  [gr-qc/0106002].

\bibitem{Horava:2009uw} 
  P.~Horava,
  ``Quantum Gravity at a Lifshitz Point,''
  Phys.\ Rev.\ D {\bf 79}, 084008 (2009)
  [arXiv:0901.3775 [hep-th]].
  
\bibitem{Nojiri:2009th} 
  S.~Nojiri and S.~D.~Odintsov,
  Phys.\ Rev.\ D {\bf 81}, 043001 (2010)
  doi:10.1103/PhysRevD.81.043001
  [arXiv:0905.4213 [hep-th]].
  
\bibitem{Cognola:2016gjy} 
  G.~Cognola, R.~Myrzakulov, L.~Sebastiani, S.~Vagnozzi and S.~Zerbini,
  arXiv:1601.00102 [gr-qc].

\bibitem{Baez:1999sr} 
  J.~C.~Baez,
  ``An Introduction to spin foam models of quantum gravity and BF theory,''
  Lect.\ Notes Phys.\  {\bf 543}, 25 (2000)
  [gr-qc/9905087].

\bibitem{Itzykson:1980rh} 
  C.~Itzykson and J.~B.~Zuber,
  ``Quantum Field Theory,''
  New York, Usa: Mcgraw-hill (1980) 705 P.(International Series In Pure and Applied Physics)

\bibitem{Cheng:1985bj} 
  T.~P.~Cheng and L.~F.~Li,
  Oxford, Uk: Clarendon ( 1984) 536 P. ( Oxford Science Publications)

\bibitem{Agashe:2014kda} 
  K.~A.~Olive {\it et al.} [Particle Data Group Collaboration],
  Chin.\ Phys.\ C {\bf 38}, 090001 (2014).
  doi:10.1088/1674-1137/38/9/090001

\bibitem{Utiyama:1956sy} 
  R.~Utiyama,
  Phys.\ Rev.\  {\bf 101}, 1597 (1956).
  
\bibitem{Kibble:1961ba} 
  T.~W.~B.~Kibble,
  J.\ Math.\ Phys.\  {\bf 2}, 212 (1961).
 
\bibitem{Sciama:1964wt} 
  D.~W.~Sciama,
  Rev.\ Mod.\ Phys.\  {\bf 36}, 463 (1964)
  [Erratum-ibid.\  {\bf 36}, 1103 (1964)].

\bibitem{Trautman:1981fd}
  A.~Trautman, ``Fiber Bundles, Gauge Fields, And Gravitation,'' {\it  In *Held.A.(Ed.): General Relativity and Gravitation, Vol.1*, 287-308 (1980).}

\bibitem{Mardones:1990qc} 
  A.~Mardones and J.~Zanelli,
  Class.\ Quant.\ Grav.\  {\bf 8}, 1545 (1991).

\bibitem{Sobreiro:2011hb} 
  R.~F.~Sobreiro, A.~A.~Tomaz and V.~J.~V.~Otoya,
  Eur.\ Phys.\ J.\ C {\bf 72}, 1991 (2012)
  [arXiv:1109.0016 [hep-th]].

\bibitem{Gribov:1977wm} 
  V.~N.~Gribov,
  Nucl.\ Phys.\ B {\bf 139}, 1 (1978).

\bibitem{Singer:1978dk} 
  I.~M.~Singer,
  Commun.\ Math.\ Phys.\  {\bf 60}, 7 (1978).

\bibitem{Zwanziger1989}
 D.~Zwanziger,
  ``{Action from the Gribov horizon},''
 Nucl. Phys.{\bf B321} (1989) 591.

\bibitem{Zwanziger:1989mf}
 D.~Zwanziger,
  Nucl. Phys.{\bf B323} (1989)513--544.
  
\bibitem{Sobreiro:2005ec} 
  R.~F.~Sobreiro and S.~P.~Sorella,
  ``Introduction to the Gribov ambiguities in Euclidean Yang-Mills theories,''
  Lectures given by SPS at the 13th Jorge Andre Swieca Summer School on Particles and Fields, Campos de Jord\~ao, Brazil, 9-22 January 2005.\\
  hep-th/0504095.
 
\bibitem{Dudal:2005na} 
  D.~Dudal, R.~F.~Sobreiro, S.~P.~Sorella and H.~Verschelde,
  Phys.\ Rev.\ D {\bf 72}, 014016 (2005)
  [hep-th/0502183].

\bibitem{Baulieu:2008fy} 
  L.~Baulieu and S.~P.~Sorella,
  Phys.\ Lett.\ B {\bf 671}, 481 (2009)
  [arXiv:0808.1356 [hep-th]].

\bibitem{Baulieu:2009xr} 
  L.~Baulieu, M.~A.~L.~Capri, A.~J.~Gomez, V.~E.~R.~Lemes, R.~F.~Sobreiro and S.~P.~Sorella,
  Eur.\ Phys.\ J.\ C {\bf 66}, 451 (2010)
  [arXiv:0901.3158 [hep-th]].
  
\bibitem{Dudal:2010cd} 
  D.~Dudal, M.~S.~Guimaraes and S.~P.~Sorella,
  Phys.\ Rev.\ Lett.\  {\bf 106}, 062003 (2011)
  [arXiv:1010.3638 [hep-th]].

\bibitem{Dudal:2011gd} 
  D.~Dudal, S.~P.~Sorella and N.~Vandersickel,
  Phys.\ Rev.\ D {\bf 84}, 065039 (2011)
  
\bibitem{Capri:2012hh} 
  M.~A.~L.~Capri, D.~Dudal, M.~S.~Guimaraes, L.~F.~Palhares and S.~P.~Sorella,
  arXiv:1208.5676 [hep-th].
   
\bibitem{Dudal:2012sb} 
  D.~Dudal and S.~P.~Sorella,
  Phys.\ Rev.\ D {\bf 86}, 045005 (2012)
  [arXiv:1205.3934 [hep-th]].

\bibitem{Pereira:2013aza} 
  A.~D.~Pereira and R.~F.~Sobreiro,
  Eur.\ Phys.\ J.\ C {\bf 73}, 2584 (2013)
  [arXiv:1308.4159 [hep-th]].

\bibitem{Dudal:2013vha} 
  D.~Dudal, M.~S.~Guimaraes, L.~F.~Palhares and S.~P.~Sorella,
  ``From QCD to a dynamical quark model: construction and some meson spectroscopy,''
  arXiv:1303.7134 [hep-ph].

\bibitem{Capri:2013naa} 
  M.~A.~L.~Capri, D.~Dudal, M.~S.~Guimaraes, I.~F.~Justo, L.~F.~Palhares and S.~P.~Sorella,
  Annals Phys.\  {\bf 339}, 344 (2013)
  doi:10.1016/j.aop.2013.09.006
  [arXiv:1306.3122 [hep-th]].

\bibitem{Capri:2015pja} 
  M.~A.~L.~Capri, A.~D.~Pereira, R.~F.~Sobreiro and S.~P.~Sorella,
  Eur.\ Phys.\ J.\ C {\bf 75}, no. 10, 479 (2015)
  doi:10.1140/epjc/s10052-015-3707-z
  [arXiv:1505.05467 [hep-th]].

\bibitem{Capri:2015ixa} 
  M.~A.~L.~Capri {\it et al.},
  Phys.\ Rev.\ D {\bf 92}, no. 4, 045039 (2015)
  doi:10.1103/PhysRevD.92.045039
  [arXiv:1506.06995 [hep-th]].

\bibitem{Capri:2015nzw} 
  M.~A.~L.~Capri {\it et al.},
  Phys.\ Rev.\ D {\bf 93}, no. 6, 065019 (2016)
  doi:10.1103/PhysRevD.93.065019
  [arXiv:1512.05833 [hep-th]].

\bibitem{Capri:2016aqq} 
  M.~A.~L.~Capri {\it et al.},
  arXiv:1605.02610 [hep-th].
  
\bibitem{Pereira:2016fpn} 
  A.~D.~Pereira, R.~F.~Sobreiro and S.~P.~Sorella,
  arXiv:1605.09747 [hep-th].
  
\bibitem{Capri:2016aif} 
  M.~A.~L.~Capri, D.~Fiorentini, A.~D.~Pereira, R.~F.~Sobreiro, S.~P.~Sorella and R.~C.~Terin,
  arXiv:1607.07912 [hep-th].

\bibitem{Inonu:1953sp} 
  E.~Inonu and E.~P.~Wigner,
  Proc.\ Nat.\ Acad.\ Sci.\  {\bf 39}, 510 (1953).
  
\bibitem{Sobreiro:2012iv} 
  R.~F.~Sobreiro, A.~A.~Tomaz and V.~J.~Vasquez Otoya,
  J.\ Phys.\ Conf.\ Ser.\  {\bf 453}, 012014 (2013)
  doi:10.1088/1742-6596/453/1/012014
  [arXiv:1211.5993 [hep-th]].
  
\bibitem{Sobreiro:2012dp} 
  R.~F.~Sobreiro, A.~A.~Tomaz and V.~J.~V.~Otoya,
  ``Gauge theories and gravity,''
  PoS ICMP {\bf 2012}, 019 (2012)
  [arXiv:1210.8446 [hep-th]].

\bibitem{Weinberg:1988cp} 
  S.~Weinberg,
  Rev.\ Mod.\ Phys.\  {\bf 61}, 1 (1989).

\bibitem{Shapiro:2006qx} 
  I.~L.~Shapiro and J.~Sola,
  J.\ Phys.\ A {\bf 40}, 6583 (2007)
  doi:10.1088/1751-8113/40/25/S03
  [gr-qc/0611055].

\bibitem{Shapiro:2009dh} 
  I.~L.~Shapiro and J.~Sola,
  Phys.\ Lett.\ B {\bf 682}, 105 (2009)
  [arXiv:0910.4925 [hep-th]].

\bibitem{Assimos:2013eua} 
  T.~S.~Assimos, A.~D.~Pereira, T.~R.~S.~Santos, R.~F.~Sobreiro, A.~A.~Tomaz and V.~J.~V.~Otoya,
  arXiv:1305.1468 [hep-th].
  
\bibitem{Piguet:1995er} 
  O.~Piguet and S.~P.~Sorella,
  ``Algebraic renormalization: Perturbative renormalization, symmetries and anomalies,''
  Lect.\ Notes Phys.\ M {\bf 28}, 1 (1995).

\bibitem{Gross:1973id} 
  D.~J.~Gross and F.~Wilczek,
  Phys.\ Rev.\ Lett.\  {\bf 30}, 1343 (1973).
  
\bibitem{Politzer:1973fx} 
  H.~D.~Politzer,
  Phys.\ Rev.\ Lett.\  {\bf 30}, 1346 (1973).

\bibitem{Zwanziger:1992qr} 
  D.~Zwanziger,
  Nucl.\ Phys.\ B {\bf 399}, 477 (1993).
   
\bibitem{Maggiore:1993wq} 
  N.~Maggiore and M.~Schaden,
  Phys.\ Rev.\ D {\bf 50}, 6616 (1994)
  [hep-th/9310111].
  
\bibitem{Bloch:2003sk} 
  J.~C.~R.~Bloch, A.~Cucchieri, K.~Langfeld and T.~Mendes,
  Nucl.\ Phys.\ B {\bf 687}, 76 (2004)
  [hep-lat/0312036].

\bibitem{Cucchieri:2006xi} 
  A.~Cucchieri and T.~Mendes,
  Braz.\ J.\ Phys.\  {\bf 37}, 484 (2007)
  [hep-ph/0605224].
  
\bibitem{Cucchieri:2009zt} 
  A.~Cucchieri and T.~Mendes,
  Phys.\ Rev.\ D {\bf 81}, 016005 (2010)
  [arXiv:0904.4033 [hep-lat]].

\bibitem{Cucchieri:2012cb} 
  A.~Cucchieri, D.~Dudal and N.~Vandersickel,
  Phys.\ Rev.\ D {\bf 85}, 085025 (2012)
  doi:10.1103/PhysRevD.85.085025
  [arXiv:1202.1912 [hep-th]].

\bibitem{Cucchieri:2014via} 
  A.~Cucchieri, D.~Dudal, T.~Mendes and N.~Vandersickel,
  Phys.\ Rev.\ D {\bf 90}, no. 5, 051501 (2014)
  doi:10.1103/PhysRevD.90.051501
  [arXiv:1405.1547 [hep-lat]].

\bibitem{Cucchieri:2016jwg} 
  A.~Cucchieri, D.~Dudal, T.~Mendes and N.~Vandersickel,
  Phys.\ Rev.\ D {\bf 93}, no. 9, 094513 (2016)
  doi:10.1103/PhysRevD.93.094513
  [arXiv:1602.01646 [hep-lat]].
    
\bibitem{Nogueira:1991ex} 
  P.~Nogueira,
  J.\ Comput.\ Phys.\  {\bf 105}, 279 (1993).
  doi:10.1006/jcph.1993.1074
  
\bibitem{Ford:2009ar} 
  F.~R.~Ford and J.~A.~Gracey,
  J.\ Phys.\ A {\bf 42}, 325402 (2009)
  [Erratum-ibid.\  {\bf 43}, 229802 (2010)]
  [arXiv:0906.3222 [hep-th]].

\bibitem{Kuipers:2012rf} 
  J.~Kuipers, T.~Ueda, J.~A.~M.~Vermaseren and J.~Vollinga,
  Comput.\ Phys.\ Commun.\  {\bf 184}, 1453 (2013)
  doi:10.1016/j.cpc.2012.12.028
  [arXiv:1203.6543 [cs.SC]].

\bibitem{Kuipers:2013pba} 
  J.~Kuipers, T.~Ueda and J.~A.~M.~Vermaseren,
  Comput.\ Phys.\ Commun.\  {\bf 189}, 1 (2015)
  doi:10.1016/j.cpc.2014.08.008
  [arXiv:1310.7007 [cs.SC]].

\bibitem{Ueda:2014sya} 
  T.~Ueda and J.~Vermaseren,
  J.\ Phys.\ Conf.\ Ser.\  {\bf 523}, 012047 (2014).
  doi:10.1088/1742-6596/523/1/012047

\bibitem{Steinhauser:2015wqa} 
  M.~Steinhauser, T.~Ueda and J.~A.~M.~Vermaseren,
  Nucl.\ Part.\ Phys.\ Proc.\  {\bf 261-262}, 45 (2015)
  doi:10.1016/j.nuclphysbps.2015.03.006
  [arXiv:1501.07119 [hep-ph]].

\bibitem{Collins:1984xc} 
  J.~C.~Collins,
  Cambridge, Uk: Univ. Pr. ( 1984) 380p
  
\bibitem{Sorella:2009vt} 
  S.~P.~Sorella,
  Phys.\ Rev.\ D {\bf 80}, 025013 (2009)
  [arXiv:0905.1010 [hep-th]].

\bibitem{Dudal:2009bf} 
  D.~Dudal, N.~Vandersickel, H.~Verschelde and S.~P.~Sorella,
  PoS QCD {\bf -TNT09}, 012 (2009)
  [arXiv:0911.0082 [hep-th]].
  
\bibitem{Faddeev:1967fc} 
  L.~D.~Faddeev and V.~N.~Popov,
  Phys.\ Lett.\ B {\bf 25}, 29 (1967).


\end{thebibliography}
\end{document}